\documentstyle[aps,prl,multicol,epsf]{revtex} 
\def\frac#1#2{{\textstyle{#1\over#2}}} 

\title{\rightline{ \small CERN-TH/2000-029 \normalsize} \rightline{ 
\small DAMTP-2000-105 \normalsize} 
\rightline{\small hep-ph/0009204 \normalsize} 
\centerline{A critical look at risk assessments for global catastrophes} } 

\author{ Adrian Kent} 
\address{ 
Department of Applied Mathematics and 
Theoretical Physics, Centre for Mathematical Sciences,\\ 
University of Cambridge, Wilberforce Road, Cambridge CB3 0WA, U.K.\\ 
} 

\date{April 2003; corrections and note added July 2015} 
\begin{document} 
\maketitle 
\begin{abstract}
  Recent papers by Busza et al. (BJSW) and Dar et al. (DDH) argue that
  astrophysical data can be used to establish small bounds on the risk
  of a ``killer strangelet'' catastrophe scenario in the RHIC and ALICE
  collider experiments.  The case for the safety of the experiments
  set out by BJSW does not rely solely on these bounds, but on theoretical
  arguments, which BJSW find sufficiently compelling to firmly exclude
  any possibility of catastrophe.
  
  Nonetheless, DDH and other commentators (initially including BJSW) 
  suggested that these empirical bounds alone do give sufficient
  reassurance.  This seems unsupportable when
  the bounds are expressed in terms of expectation value --- a good
  measure, according to standard risk analysis arguments.  For
  example, DDH's main bound, $p_{\rm catastrophe} < 2 \times 10^{-8}$,
  implies only that the expectation value of the number of deaths is
  bounded by $120$; BJSW's most conservative bound implies the
  expectation value of the number of deaths is bounded by $60000$.  
  
  This paper reappraises the DDH and BJSW risk bounds by comparing risk
  policy in other areas.  For example, it is noted that, 
  even if highly risk tolerant assumptions are made and no value is
  placed on the lives of future generations, a catastrophe
  risk no higher than $\approx 10^{-15}$ per year would be required 
  for consistency with established policy for radiation hazard risk
  minimization.  Allowing for risk aversion and for future lives, a
  respectable case can be made for requiring a bound many orders 
  of magnitude smaller.
  
  In summary, the costs of small risks of catastrophe have been
  significantly underestimated by BJSW (initially), by DDH and by other 
  commentators.  Future policy on catastrophe risks would be more
  rational, and more deserving of public trust, if acceptable risk
  bounds were generally agreed ahead of time and if serious research 
  on whether those bounds could indeed be guaranteed was carried out
  well in advance of any hypothetically risky experiment, with the
  relevant debates involving experts with no stake in the
  experiments under consideration.  
  
\end{abstract} 
\vskip10pt 
\pagebreak
\section{Introduction}\label{introduction} 

Speculative suggestions are occasionally made about ways in which new
physics experiments could hypothetically bring about a catastrophe
leading to the end of life on Earth.  Some of these hypothetical
catastrophes, including the ``killer strangelet'' scenario considered 
in this paper, would also lead to the destruction of the 
planet and wider catastrophic consequences.
In any case, the proposed catastrophe mechanisms generally rely
on speculation about hypothetical phenomena for which there
is no evidence, but which at first sight do not contradict the 
known laws of physics.  
Sometimes, such pessimistic hypotheses can be countered by arguments which
show that the existence of the catastrophe mechanism is highly
improbable, either because closer analysis shows that 
the proposed mechanism does in fact contradict well established 
physical principles, or because its existence would imply effects 
which we should almost certainly have observed but have not.  

Unfortunately, there is a 
difficulty in making an argument of this type sufficiently 
reassuring.  One would like to be reassured that the chances
of inadvertently triggering a global catastrophe are very small
indeed before going ahead with an experiment.  But finding
arguments which justify this conclusion with the appropriate
level of confidence may be very hard, if not impossible.  
Discouragingly few attempts to grapple with this issue have been made.  
In fact, even the obvious and fundamental question --- 
how improbable does a catastrophe have to be to justify
proceeding with an experiment? --- seems never to 
have been seriously examined.  The aim of this 
paper is to face this question squarely, in the hope
of stimulating further debate.

The particular stimulus for this paper was the debate over
the safety of the RHIC supercollider experiments now underway
at Brookhaven, and the ALICE experiments proposed by CERN. 
Speculation about possible disaster scenarios in these
experiments led to some pressure for the experiments to 
be deferred or cancelled.  In response, reports and papers
were written that were used to justify commencing the RHIC
experiments on the grounds that, inter alia, ``Cosmic
ray collisions provide ample reassurance that we are
safe from a $\ldots$ catastrophe at RHIC''\cite{bjswone}
and ``Beyond reasonable doubt, heavy-ion experiments at RHIC
will not endanger our planet.''\cite{ddh}.
Here I will contend that the risk bounds obtained are actually
{\it not} small, taking into account the scale of the catastrophe,
either according to standard risk analysis or when 
compared with other adopted standards for acceptable
risk to the public.  Since the criteria used by Brookhaven to justify
proceeding were developed by theoretical physicists and administrators,
not by broader representatives of the public or by professionals
in risk management, it seems desirable to bring these issues before a wider
audience for the purposes of informed discussion and the 
formation of sounder public policy in future.  

\section{Historical Examples} 

The first catastrophe mechanism seriously considered seems to have been
the possibility, raised in the 1940s at Los Alamos 
before the first atomic bomb tests, that 
fission or fusion bombs might ignite the atmosphere or oceans 
in an unstoppable chain reaction.   Investigation led to an  
analysis by Konopinski et al.\cite{kmt} which fairly definitively refuted 
the possibility.  
Compton was later reported, in a published interview\cite{buck} with
Pearl Buck, as saying that he 
had decided not to proceed with the bomb tests if it were proved 
that the chances of global catastrophe were greater than 
three in a million, but that in the event calculation proved
the figures slightly less.  

It is hard to understand how any meaningful calculation 
could have produced such a risk figure. The analysis of 
Ref. \cite{kmt} gives convincing arguments against the
possibility of a catastrophic chain reaction, based on well established
physical principles.  It concludes that it is unreasonable to expect
a chain reaction propagated by nitrogen-nitrogen fusion reactions, 
and that an unlimited chain reaction consuming the atmosphere is 
less likely still.  Other possible reactions, involving protons
in clouds of steam liberated from the oceans, are also considered
and argued to be less dangerous still.  Konopinski et al. do note 
the ``distant probability'' that 
the mode of propagation of the reaction in the atmosphere might
be more complicated than their analysis allows, in which case
its conclusions might not apply, and they suggest that the 
complexity of their argument and the absence of a satisfactory
experimental basis for it makes further work on the subject
highly desirable.  However, they offer nothing resembling a 
catastrophe risk estimate, nor any results from which a 
quantitative estimate could be derived. 

Yet, so far as I know, Compton never made an attempt to correct
Buck's account.  Had she simply misunderstood, it would have been
easy for Compton to disclaim the statement.  And, had it not 
reflected his views, he would surely have wanted both to set the
historical record straight and to defend his reputation against
the charge of unwisely gambling with the future of humanity.  
The natural inference seems to be that Compton did 
indeed make the statement reported.\footnote{
In April 2000, in an attempt 
to understand this puzzling statement of Compton's, I contacted Hans Bethe, 
a key figure in both the Los Alamos project and the theoretical work
which led to the conclusion that the possibility of an atomic bomb
explosion leading to global catastrophe was negligible.  
His view, 
relayed by an intermediary (Kurt Gottfried), was that the 
analysis of Konopinski et al. was definitive and does
not allow one to make any meaningful statement about probabilities 
since the conditions that must be met cannot be reached in any
plausible manner.   Bethe suggested that the $3 \times 10^{-6}$ figure
was made up by Compton ``off the top of his head'', and is ``far, far too
large''\cite{bethe}.}
If so, although the risk figure itself appears unjustifiable, 
Compton presumably genuinely believed that an actual risk (not
just a risk bound) of $3 \times 10^{-6}$ of global catastrophe 
was roughly at the 
borderline of acceptability, in the cause of the American and allied
effort to develop atomic weapons during World War Two.  Apparently 
the figure did not worry 1959 {\it American Weekly} readers greatly,
since no controversy ensued.  It would be interesting
to compare current opinion on the acceptability of a 
$3 \times 10^{-6}$ risk of global catastrophe, in the
circumstances of the Los Alamos project or otherwise. 

Another hypothetical catastrophe was examined some time ago by 
Hut and Rees\cite{hutrees,hut}.  They considered the possibility
that the vacuum state we live in is not the true vacuum, but 
merely a local minimum of the effective potential.  
They asked whether, if this were the case, new generations of 
collider experiments could trigger a catastrophic transition to 
the true vacuum, destroying not only the Earth but (eventually) 
all presently stable forms of matter in the cosmos.  
They showed that the probability of this occurring artificially in
present or foreseeable collider experiments is considerably
smaller than the probability of it having occurred naturally within
our past light-cone.\footnote{These results 
are reviewed in Ref. \cite{bjswtwo}.  Whether they offer
adequate reassurance that no forseeable collider experiment
will be unacceptably risky deserves reconsideration in 
the light of the arguments set out below.} 

Most recently, in response to some (rather unfocussed) 
public concern\cite{concern},
the possibility of some catastrophe arising from prospective 
experiments at the Brookhaven relativistic heavy ion collider (RHIC) 
was reviewed by Busza et al. (BJSW) and 
Dar et al. (DDH).\footnote{DDH also considered the ALICE experiments,
scheduled to take place later at CERN.  
Their bounds on the risk of catastrophe ensuing from ALICE
will not be considered here, though it is worth
noting that even DDH regard them as inadequate and 
suggest further investigation.}
Both groups paid most attention to the 
``killer strangelet'' catastrophe 
scenario, which would arise if negatively charged
metastable strange matter existed and could be produced in
the experiments. 
As well as giving theoretical arguments 
against the hypotheses involved, both groups offer risk bounds
inferred from empirical evidence.  BJSW's proposed bounds on the probability
of catastrophe during the ten year lifetime of RHIC, derived from the
survival of the Moon for $4.5$ billion years, range from $10^{-5}$ 
to $2 \times 10^{-11}$, depending on how conservative the
assumptions made are\cite{bjswtwo}.  DDH's bounds, derived from 
the observed rate of supernovae, range 
from $2 \times 10^{-6}$ (their pessimistic bound for a very slow
catastrophe, in which the Earth is prematurely destroyed at some
point in the billion years before it would anyway be consumed by the expanding
Sun) to $2 \times 10^{-8}$ (their main bound)\cite{ddh}.

As the quotations extracted in Section IV attest,
both groups originally\cite{bjswone,ddh} suggested their empirical bounds 
alone were adequate to show that the experiments were safe.  
If correct, this conclusion would obviously be particularly welcome, since 
it would remove any need to evaluate the degree of 
confidence which should be placed in the theoretical 
arguments.  The view that the empirical bounds were indeed
adequate was also expressed in
commentaries\cite{bnlweb,glashowwilson}.  

However, there are good reasons, explained below, to believe that
this conclusion is incorrect, and indeed the claim was withdrawn
by BJSW, after criticisms from the author of this paper. 
BJSW produced a second version of their preprint, removing 
the reassuring characterisations of their risk bounds
and instead disavowing any attempt to decide what is an acceptable upper
bound on $p_{\rm catastrophe}$.  In this revised version, BJSW 
also accept that the arguments for their empirically derived risk 
bounds could be invalidated if some additional pessimistic 
hypotheses were correct.\footnote{Some further possible 
loopholes in those arguments are listed in Ref. \cite{akrhicphys}.}

BJSW's revision of their preprint was 
an adequate response, from a purely scientific perspective. 
The public policy implications are troubling, however. 
My understanding is that the US government's authorisation 
for the RHIC experiments to proceed
was given partly on the basis of BJSW's original arguments\cite{bjswone}, 
whose discussions
of risk were gravely flawed, as the quotations considered in
Section IV illustrate.  
Public concern was countered by widely
publicised\cite{sciamreassurance,nsreassurance}
reassurances\cite{bnlweb,glashowwilson}
that the risk was negligible, also relying heavily on the risk appraisals
given in BJSW's original preprint\cite{bjswone}.   
As far as I am aware, no effort was
made by Brookhaven to reobtain authorisation on the basis of
BJSW's revised assessment, or to bring what is a significantly
revised case to media and public attention.\footnote{
Indeed, the Brookhaven web pages continue, in April 2003, to direct 
readers to the original unamended version of BJSW's preprint: see
http://www.bnl.gov/rhic/docs/rhicreport.pdf.} 
In my opinion, such efforts should have been made.  

It should be noted that BJSW stress\cite{bjswtwo} in the revised
version of their paper that they regard the theoretical arguments 
alone as sufficiently compelling.  This may be a defensible position, 
but it is not the case that was originally made and publicised.  

\section{Scope of this paper}

This paper is meant as a contribution to the debates over 
hypothetical catastrophe scenarios in the RHIC and ALICE collider
experiments and over other hypothetical or real global catastrophe risks.  
It focusses on the key question: 
what risk of catastrophe could be acceptable?  
Other relevant questions are not considered.  
In particular, in the case of the collider experiments, no attempt is made 
to infer quantitative risk bounds from the qualitative 
theoretical arguments against the possibility of catastrophe,
or to consider BJSW's conclusion that the theoretical arguments alone
offer sufficiently compelling reassurance\cite{bjswone,bjswtwo}.  

The interest in this debate is not, of course, purely or mainly intellectual. 
The aim is to improve future policy over catastrophe risks.  
In particular, lessons can and should be learned 
from the evident flaws in BJSW's and DDH's discussions of
risk.  
It is obviously unsatisfactory that the question
of what constitutes an acceptable catastrophe risk should continue to 
be decided, in an ad hoc way, according to the personal
risk criteria of scientists whom those in charge of experimental
facilities choose to consult. 
Those criteria, however sincerely held and thoughtfully constructed,
may be unrepresentative of general opinion or of expert opinion
in risk analysis. 

Worse still, history suggests the risk criteria actually 
used may not be at all thoughtfully constructed. 
Compton's reported opinion suggests, and the mischaracterisations of Refs. 
\cite{bjswone} and \cite{ddh} illustrate 
very clearly, that scientists whose expertise is not in risk analysis or public
policy cannot necessarily be relied on either to interpret the risk
implications of the science correctly or to consider elementary arguments 
that tend to suggest more cautious risk criteria than can easily
be satisfied.   Relying on such inexpert appraisals
is neither in the public interest nor
the long term interests of science.
Scientists and scientific institutions need to work to 
gain, maintain and justify public trust.  Arguments which 
suggest that an experiment should
proceed simply because the global catastrophe 
risk appears fairly low, without comparison to any 
pre-existing thresholds or guidelines, may not only fail
to reassure, but may (not unreasonably) be interpreted as public 
relations exercises, intended to support a prejudged 
conclusion, rather than dispassionate scientific analyses. 
As Calogero notes\cite{calogero}, this has a long term cost
for the credibility on questions of risk
not only of those directly involved, but
of all scientists, and the likely long term consequence is less
informed and more irrational public debate and public policy.  

It may well not be possible to reach a complete consensus 
on firm guidelines.  It seems unlikely, for instance, that 
some clear agreement will emerge that global catastrophe risks
are small enough to be of negligible concern if and only if 
lower than, say, $10^{-20}$.  Life is more complicated than that,
and democratic debate more multi-faceted.  Nonetheless --- indeed,
for this very reason --- it would 
be valuable to have a spectrum of carefully argued opinion in the 
literature.   I hope that the arguments below may spark
further discussion. 

\section{BJSW and DDH's risk bounds for the ``killer strangelet''
  scenario} 

We turn now to the particulars of the catastrophe risk concerns raised over the
RHIC and ALICE experiments, and specifically to the hypothetical
``killer strangelet'' catastrophe scenario analysed in some
detail by BJSW and DDH. 
The ``killer strangelet'' scenario requires: 
(i) that stable strange matter exists, (ii) that a valley of stability 
exists for {\it negatively} charged strangelets, (iii) that 
negatively charged metastable strangelets could be 
produced in the $\approx 40~{\rm TeV}$ Au-Au ion collisions 
planned at RHIC, (iv) that a strangelet so produced could survive collisions
which bring it towards rest in surrounding matter, (v) that 
it would then fuse with nuclei, producing larger  
negatively charged strangelets, in a runaway reaction which 
eventually consumes the Earth.\footnote{Aficionados of understatement
may admire BJSW's description:\cite{bjswone} ``a catastrophic process 
with profound implications for health and safety''.} 
The theoretical arguments\cite{bjswtwo,ddh}  against (i)-(iii) are
generally regarded as convincing.  
If (i)-(iii) were nonetheless true, (iv) and (v) would also be
plausible.  

If (i)-(v) were true, killer strangelets should also be 
produced in naturally occurring high energy heavy ion collisions, 
which take place when cosmic rays collide with one another or with
heavy nuclei in celestial bodies.  
Naturally produced killer strangelets would be able 
to initiate runaway reactions capable of destroying asteroids, 
satellites such as the Moon, or stars.  From the fact that 
the Moon has survived for $4.5$ billion years, and from the
fact that astronomical observations are not consistent with stars being 
converted into strange matter at any significant rate, 
bounds on the risk of catastrophe at RHIC can be 
derived\cite{bjswtwo,ddh}.  

Unfortunately, these derivations
require assumptions about the types of interaction
which produce strangelets, the velocity distribution of the 
strangelets produced, their interactions with nuclei, and their
stability\cite{bjswtwo,ddh,akrhicphys}.
For this reason, BJSW and DDH give various bounds, derived by
making more or less conservative assumptions.
Even the weakest of these requires some 
assumptions\cite{bjswtwo,akrhicphys}.

Without knowing what level of confidence we can have in the 
relevant assumptions --- a question which neither group 
addresses quantitatively --- it is difficult to see how 
the bound figures can really be meaningful\cite{akrhicphys}.
But even if the figures cannot really be justified, the comments 
made on them by BJSW, DDH 
and others give an interesting and valuable 
insight into the risk criteria of 
physicists and administrators involved in RHIC policy.

Assuming that RHIC runs for the scheduled $10$ years,
DDH obtain $p_{\rm catastrophe} < 2 \times 10^{-8}$
for a fast catastrophic destruction of the Earth and 
$p < 2 \times 10^{-6}$ for a
slow destruction that would be completed in the billion years
before the Sun expands beyond Earth orbit.  

DDH describe these results as ``a safe and stringent upper bound on the risk 
incurred in running [RHIC]''.  They add 
that the two bounds respectively imply that ``it is safe to run RHIC for 
500 million years'' and that ``running the RHIC experiments for
five million years is $\ldots$ safe''.  These last two statements are, of
course, incorrect.   DDH's bounds, if valid, would 
establish only that it would be unlikely that the Earth would be
destroyed very early in a RHIC experiment run over the relevant
periods: the bounds are consistent with a high probability of 
destroying the Earth at some point during these hypothetically
extended experiments.\footnote{That these
statements misrepresent their results was pointed out to
DDH by the author in January 2000.} 

In the first 
version\cite{bjswone} of their paper, BJSW 
described DDH's result as ``a factor of $10^8$ below the value
required for the safety of RHIC''.  This, of course, is also
incorrect: a risk bound $10^8$ times that of DDH's would be 
consistent with a high probability of destroying the Earth
within $5$ years of the RHIC experiment --- a risk level which
even the most gung-ho physicist could hardly describe as ``safety''. 
Using their own independent analyses, BJSW
derive results that imply the following bounds from the survival of the Moon, 
given various assumptions (their Cases I-III) about strangelet
production, and again assuming that RHIC runs for the scheduled
ten years:
$p_{\rm catastrophe} < \approx 2 \times 10^{-10}$, 
$p_{\rm catastrophe} < \approx 10^{-4}$, 
$p_{\rm catastrophe} < \approx 2 \times 10^{-5}$. 
They described the second and third of these cases as still
leaving ``a comfortable margin of error''.   These comments, and that
quoted at the start of this paragraph, are so obviously
inapplicable --- no sane person would seek to reassure the public
by suggesting that a risk bound of $1$ in $\approx 10000$
of destroying the Earth represented a comfortable margin of error ---
that I suspect they must reflect some
surprising confusion on BJSW's part at the time of writing
Ref. \cite{bjswone}.

BJSW refined their calculations in the second version of
their paper, extracting an extra factor of ten and producing bounds 
for a ten year run of the RHIC experiment of (Cases I-III): 
$p_{\rm catastrophe} < \approx 2 \times 10^{-11}$, 
$p_{\rm catastrophe} < \approx 10^{-5}$, 
$p_{\rm catastrophe} < \approx 2 \times 10^{-6}$. 
In this revised version, which followed criticisms of the comments noted
above, no judgement is made as to whether
any of the bounds are satisfactory.  To quote BJSW:
``We do not attempt to decide what is an acceptable upper limit
on [$p_{\rm catastrophe}$], nor do we attempt a `risk analysis', 
weighing the probability of an adverse event against the severity
of its consequences.''\cite{bjswtwo}
We use the revised bounds in the following discussion, referring
to them simply as BJSW's bounds. 

DDH's main bound --- $p_{\rm catastrophe} < 2 \times 10^{-8}$ over
the $10$ year life of RHIC --- has been widely 
referred to\cite{bjswone,ddh,glashowwilson,bnlweb}   in terms
which suggest that it alone would be sufficiently reassuring to 
require no further analysis or risk 
optimisation.
My impression is that many numerate and thoughtful people would 
disagree.   My own reasons for doing so are given below. 

\section{Risk bounds and risk estimates: an important caveat} 

It is important to stress that DDH's and BJSW's empirical arguments 
produce {\it bounds} on the risk of catastrophe, not {\it estimates} of that 
risk.\footnote{In contrast, Compton's reported statement on the risk
of destroying life on Earth by a fission explosion is given in the
form of a risk estimate --- though, as noted above, it was not justifiable.}  
Their bounds are based on the fact that we do not observe something that
we should expect to observe if the risk were larger than
some value $p$.  A negative result of this form tells us nothing about the
actual value of the risk.  Everything in DDH's and BJSW's analyses is
consistent with the true risk of catastrophe being zero --- and 
if current theoretical understanding is correct, the risk 
is indeed precisely zero. 

When the destruction of the Earth is in question, though, 
it would be preferable not to 
have to rely on theoretical expectations alone.  
As Glashow and Wilson put it\cite{glashowwilson},
``The word `unlikely', however many 
times it is repeated, just isn't enough to assuage our fears of 
this total disaster.''  
Hence the interest in looking at naturally occurring versions of
the experiment, verifying that they have not resulted in catastrophes,
and so producing firm bounds on the risk of catastrophe.  

Unfortunately, this approach has its pitfalls and limitations.  
Comparing the effects expected 
from hypothetical killer strangelets produced in naturally 
occurring heavy ion collisions and at RHIC
is not completely straightforward.  Theoretical 
assumptions need to be made in order to derive risk bounds.  
Unless we are very confident indeed in those assumptions, 
we cannot validly infer very small risk bounds 
this way\cite{akrhicphys,bjswtwo}.
And in any case, Nature may not necessarily have 
done versions of the experiment we are interested in
often enough to produce sufficiently strong risk bounds.   

How do we begin to decide what constitutes a sufficiently
strong risk bound?  It seems to me that 
the correct approach in appraising risk bounds
is to make worst case assumptions.  
So, if we are assured that $p_{\rm catastrophe} \leq p_0$, and 
we have to decide whether that bound alone is sufficient
reassurance, we have to ask whether we would be happy
to proceed if we knew that $p_{\rm catastrophe} = p_0$. 
If not, then the bound alone is not sufficiently reassuring. 

Such a bound might still form part of a compelling case
for the safety of an experiment if it could be combined with 
other arguments.  For instance, in the case of RHIC,
it might be argued that a combination of the theoretical
arguments and empirical bounds is sufficiently reassuring,
even if neither would be alone.\footnote{Arguments along these lines 
have been suggested in informal discussions, but to the best of
my knowledge none has been set out in print.  
Such an argument would need to be made very carefully, since the 
theoretical arguments and empirical bounds are not independent. 
As already noted, the empirical bounds still rely on
theoretical assumptions, and if theoretical expectations
were incorrect, the derivation of the empirical bounds might
also be affected.} 

I will not consider such arguments here.
Nor --- to reiterate --- do I examine whether the 
theoretical arguments alone are sufficiently reassuring.
The discussion below considers only the narrow question of 
whether the empirical bounds alone would suffice.  

\section{Risk versus expectation}\label{risksec} 

The destruction of the Earth would entail the 
death of the $\approx 6 \times 10^9$ human population
and of all other species, the loss of the historical record
of the evolution of its biosphere, and the loss
of almost all record of the culture developed by 
humanity.\footnote{A few spacecraft, including the
message-bearing Pioneer and Voyager craft, would survive,
as would the -- continually attenuating -- electromagnetic
signals generated on Earth.}
Added to these are the opportunity costs arising from the 
absence of future generations. 

Consider for the moment just the number of human deaths.  
If an experiment were expected to
cause one human death, in the everyday use of the term --- that is,
it was likely that at least one person would die as a result of
the experiment --- its health and safety implications could not
be said to be negligible.  Now, when we are dealing with small
risks of large catastrophes, we cannot directly use this measure:
an experiment with small risk is expected to cause no human deaths,
in the sense that this is the likeliest outcome.  
However, we can calculate a related measure: the statistical
expectation value of the number of 
human deaths.  The expectation value of the number of human deaths
ensuing from an Earth-destroying 
catastrophe is $ E_d = p_{\rm catastrophe} N$, 
where $N \approx 6 \times 10^9$ is the current human population.  

So, {\it if} we accept $E_d$ as an appropriate measure of the 
seriousness of a risk --- and the next section explains why we should ---
then any risk that does 
not satisfy
\begin{equation}\label{negl}
p_{\rm catastrophe} \ll  1.6 \times 10^{-10} 
\end{equation} 
is not negligible.   

Of the bounds above, neither of DDH's ensure that (\ref{negl})
is satisfied, nor do the second and third of BJSW's.  BJSW's least
conservative lunar survival bound (Case I) comes closer, but still
fails unless a factor of $1/8$ --- i.e. in this case a 
probability of $1/8$ of causing one human death --- is regarded
as negligible.  

Making the comparison in terms of expectations, DDH's main (tighter)
bound implies that the expectation value of the
number of human fatalities caused by 
RHIC over ten years will not exceed $120$. 
Put this way, this bound seems far from adequately reassuring. 

\section{Is the expectation value of the number of fatalities relevant?} 

It might perhaps be argued that the preceding calculation is 
misleading.  After all, DDH's bounds on $p_{\rm catastrophe}$
represent probabilities small enough to be negligible in most circumstances.
Most of us take $2 \times 10^{-8}$ risks of death in our stride: the
risk of a typical US citizen dying in a shark attack in any given
year is comparable.   
Translating the bound value into $E_d$, the expected number of
fatalities, makes it seem significant. 
But is it really reasonable to use expected fatalities as a measure of
the safety of a risk bound?  

Actually, the next section argues that considering $E_d$ alone still 
greatly {\it underestimates} the cost.  
But first let us consider whether requiring $E_d < 1 $ gives a 
sensible upper bound on negligible risk, assuming 
that it is agreed
that the certainty of causing one death would not be negligible.
I believe most expert opinion would agree that it does, for
the following reasons.

First, everyone agrees that in carrying out any risk analysis
we need the cost or utility of the various outcomes, not merely
their probability: a $10^{-3}$ chance of losing one dollar is 
better than a $10^{-3}$ chance of losing one million dollars, and 
so on.  Second, a fundamental principle of 
risk analysis is that in normal circumstances rational people
are risk averse.  If $X$ represents a random process whose
possible outcomes $x_i$ have probabilities $p_i$, and if $V(x_i )$ 
represents the value to the community of outcome $x_i$, then the 
value $V(X)$ of a single run of $X$ --- that is, the value of 
allowing precisely one of the $x_i$ to happen, with 
respective chances $p_i$ --- is generally assumed to obey
\begin{equation}
V (X ) \leq \sum_i p_i V (x_i )  \, . 
\end{equation} 
The values of undesirable outcomes, of course, are negative:  
we refer to $- V( x_i ) $ as the cost of $x_i$. 
A second principle is that the utility or cost function is
concave.  Applied to the cost of a loss of human lives,
this implies that the cost to society of $N$ deaths is at least $N$
times as great as the cost of $1$ death: if $x_N$ and $x_1$ represent
the two events, then 
\begin{equation}
V (x_N ) \leq N V( x_1 )
\end{equation}  

The principles of risk aversion and concave utility explain, for example, 
why it can often be rational to take out insurance, even
though on average the insurance company expects to make a profit 
and the customer a loss.  Similarly, it explains why
investors almost universally require investments that involve higher
risk to offer a higher expected profit in compensation.   
By considering a random process $X$ with probability $2 \times 10^{-8}$ 
of killing $6 \times 10^9$ people and probability $ ( 1 - 2 \times
10^{-8} )$ of killing no one, we see these principles together
imply that a $2 \times 10^{-8}$ chance of killing 
$6 \times 10^9$ people is at least as bad as the certainty of 
killing $120$ people.  

In summary, to demonstrate that the risk is negligible, we would
need to show that $E_d$ is considerably smaller than one --- 
precisely how much smaller depending on precisely how risk averse
one is when it comes to global catastrophe. 
Neither DDH's nor BJSW's bounds satisfy this criterion.  
To speak of the bounds being ``safe and stringent''
or guaranteeing ``comfortable margins of error'' is, on 
this analysis, simply incorrect. 
Similarly, to demonstrate that the risk is acceptable, it would
be necessary (though not necessarily sufficient) to show that $E_d$ 
is small enough that the certainty of the experiment
killing $E_d$ people would be acceptable.  
Put another way, if it would be unacceptable for the experiment
to lead to the certain loss of $E_d$ lives, then a risk at the
bound value would be unacceptable.  

Suppose, counterfactually, that we knew that the RHIC
experiment were certain to kill precisely $N$
people (and no more).  What value of $N$ would be acceptable?  
Answers will vary, but my guess is that most would be somewhere
in the range $<10$ or so.  In particular, I think it clear that 
RHIC would not have obtained political authorisation if it
was thought certain to kill precisely $120$ people: that would
be regarded as an unacceptably high cost.  From the discussion
of this section, it follows that a global catastrophe risk at the DDH bound
value would be at least as unacceptable.  

Although the observations made in this section are elementary, it 
is worth noting that the CERN panel did not acknowledge their
validity.  The response of Alvaro de Rujula, the panel leader,
is accurately summarised by his opinion, quoted in New
Scientist\cite{drns}, that it 
is ``absurd'' to take the risk bound probability and multiply it
by the global population.  I recommend contemplation of this
comment to anyone inclined to automatic faith in the risk
analysis expertise of scientists chosen by institutions
to argue for the safety of their experiments.  

\section{Comparison with existing risk policies}  

DDH's and BJSW's risk bounds 
were presented and discussed by DDH\cite{ddh} and
BJSW\cite{bjswone}, in a statement by John Marburger, then director of
Brookhaven\cite{bnlweb}, and in a commentary by Glashow and 
Wilson\cite{glashowwilson}. 
None of these discussions make comparisons with risk 
criteria or optimisation procedures 
applied to other potentially hazardous activities.  
This is unfortunate, since risk comparisons are generally
illuminating, and in this case suggest that regarding the risk 
bounds {\it per se} as adequate would be wildly 
inconsistent with at least some established policy.  

For example, the UK National Radiological Protection Board requires
that the risk of serious deleterious health effects arising from
a nuclear solid waste disposal facility must always be bounded
at below $10^{-5}$ per year, that risk optimisation procedures should
be continued until the risk is below $10^{-6}$ per year, and that
the risk of low probability natural events which could lead to 
serious deterministic health effects should be separately bounded
at $10^{-6}$ per year\cite{nrpb}. 
The risk figures apply to the critical 
group of individuals, typically numbering between a few and a few 
hundred, whose habits or location render them most at risk.
Quite typically, the events whose risk is bounded 
would be expected to kill fewer than $10$ people.

In summary, according to established policy for these 
radiation hazards, it is not acceptable to incur a risk of 
greater than $10^{-6}$ per year of killing $\approx 5$ people.  The risk
aversion arguments of the last section suggest that 
a consistent policy on catastrophe risk should treat a risk 
greater than $10^{-15}$ per year of killing the global population as 
even less acceptable.  An acceptable 
risk bound should thus imply 
\begin{equation}
p_{\rm catastrophe} \ll  10^{-15} {\rm~per~year}  \, . 
\end{equation}

\section{Future lives} 

The discussion so far has considered only the expected cost due
to immediate human fatalities, neglecting the other costs
mentioned earlier.  These are very hard to quantify in any
commensurable way.
(What is the value of the rest of the biosphere compared to 
that of the human population?  What price do we put on the 
historical record?)  However, it is, at least arguably, possible to assign 
a sensible and commensurable value to one of these further costs ---
the loss of future generations --- by estimating the number of 
future human lives which would not take place 
if the planet were destroyed in the near future.  

This line of argument cannot be followed without addressing
two rather complex questions: Should we value our 
successors' lives as highly as those of our contemporaries?  
And can we say anything meaningful about the likely fate of humanity over 
the billion years of life the Earth has left (or beyond)?   

To the first, my own answer is ``yes'', partly because I
cannot see any good reason to prefer an unknown contemporary to 
an unknown successor, and partly because it seems to me our lives
have value in the first place largely because 
they form part of ongoing human history.  
This view finds some support in established policy: the
UK National Radiological Protection Board guidlines cited
above also explicitly state that those living at any time in
the future should be given a level of protection at least
equivalent to that given to those alive now.  

As for futurology, there are obviously so many
unknowns that attempting detailed analysis seems pointless.  
I offer only a crude calculation, which is 
obviously open to criticism, but at least suggests a starting
point for discussion.  

Suppose, optimistically, that humanity has a reasonable chance of 
surviving (in some form) for the lifetime of the Earth.
Suppose also that there is a reasonable chance of 
arranging things so that the sum global 
quality of life is at least at the level of today, and 
the global population is roughly today's: $10^{10}$ in round figures.
And suppose we neglect the possibility that the human 
lifespan may increase beyond $10^2$, on the grounds that it is
arguably irrelevant: arguably, one can make a reasonable approximation
--- reasonable, that is, given the uncertainties in the entire
discussion --- by considering
the total number of person-years, so that for instance a $700$-year life
is equated to seven $100$-year lives.  Let us also, conservatively, 
neglect the effect of migration to other planets some time in 
the future, which would presumably (i) allow the human population to
vastly increase over the next billion
years and (ii) permit humanity to survive beyond a billion years.  

The cost of the destruction of the Earth today would then be 
roughly $10^{10} 10^{-2} 10^9 = 10^{17}$ lives, or $10^7$ greater 
than the cost earlier calculated.  
Including this factor in the earlier calculation derived from
the NRPB's risk bounds would mean that an acceptable risk bound should  
imply 
\begin{equation}
p_{\rm catastrophe} \ll 10^{-22} {\rm~per~year~} \, .
\end{equation} 

This further proposed tightening of risk bounds is controversial
in all sorts of ways.   To mention just one: regarding 
the loss of future lives as a separate cost raises the
question --- a cost to whom?  To those potential future generations,
deprived of the possibility of existence?  To us, deprived of 
descendants and successors?  Both, I think --- but
I concede that both lines of thought raise some difficulties.  

Another way of approaching the question is to ask a different
hypothetical question: would it be far worse if all of us
(or all life on Earth) were killed than if almost all of us
(or almost all life on Earth) were, supposing that in the 
second case the planet remained otherwise fit for life? 
Answering ``yes'' suggests placing a high relative value on 
future lives, since the number of immediate fatalities is
almost the same in both cases.  Those who answer ``no'' will
presumably not find any of the arguments of this section
convincing.  My impression is that the question has not 
been widely enough debated for it to be possible to say which view
(if either) reflects general opinion.   For the moment, then,
whichever view one holds on future lives, it is worth 
bearing in mind that the majority view, on which
catastrophe risk policy should properly be based, 
may turn out to differ.  
  
\section{Some counterarguments}  

Calculations and comparisons are indispensable in 
rationalising risk policy and in highlighting 
inconsistencies.  However, there is no generally agreed
set of principles from which we can decide policy in every
instance.  Politics do not form a subset of mathematics. 
The above arguments could well be opposed on many different 
grounds.  I consider here some counterarguments 
which have been suggested to me in discussions.  

\begin{itemize}

\item

One obvious criticism is that the arguments above consider
the cost of a catastrophe risk but not the benefits gained
by taking the risk.  For that reason, it may be argued,
they are bound to produce over-cautious prescriptions.
After all, no risk at all is worth taking unless there is
some benefit.  Without a cost-benefit analysis, no sensible
conclusion can be reached.  

This is a partially fair criticism, but only partially.  
It should be stressed that it does not apply to the 
argument of section \ref{risksec}, since using the number of deaths 
as a measure of safety is justified by comparing the implicit 
cost-benefit tradeoffs in conclusions that
would be generally agreed.  It seems to me pretty uncontroversial
that, despite the benefits of RHIC, the 
experiment would not be allowed to proceed if it
were certain (say, because of some radiation hazard) to 
cause precisely $120$ deaths among the population at large. 
If that is accepted, it follows that a risk 
at the DDH bound value would be unacceptably high, {\it even 
when} the benefits of RHIC were taken into account.

That said, some forms of cost-benefit analysis might indeed
suggest that requiring $p_{\rm catastrophe} \ll 10^{-15}$ per year
or lower may be over-stringent.  
The likely immediate benefit of the RHIC experiments
--- advancing our understanding of basic science --- is not
negligible.   Moreover, the possible benefits presumably 
include at least some probability --- perhaps small, but not
necessarily small compared to $10^{-15}$ --- of contributing in some
presently unforeseen way to a discovery with a very large beneficial
impact on future human lives.  The foresight problem is particularly
acute here, of course, since one can also imagine low probability
outcomes, other than the catastrophe scenario, with a large
negative impact.  But, if one takes the view that scientific and
technological progress have on balance been beneficial 
and are likely to continue to be, the small possibility of a
benefit that would save (or enable) many future human lives gives, 
at least in principle, something 
to offset against the small possibility of a catastrophe.  

It is important to be clear, though, that by definition no 
cost-benefit argument could justify a claim that the risks involved
are negligible.  Rather, it would
make the case for proceeding with RHIC by suggesting that the 
risks, though possibly not negligible, were justified by the benefits.  
This is not the case which has been made.  Such a case 
might or might not be widely accepted. 

\item 

It may be argued that the more stringent risk criteria 
suggested above for global catastrophe, even if rationally 
justifiable in theory, are impossibly 
Utopian.  If we took them seriously, and attempted to ensure that 
they were satisfied before proceeding with any enterprise, we would 
stop, not only collider experiments, but many other human activities. 
Progress would become impossible; life might be made unliveable.  

Maybe --- but I would be cautious about accepting this sort of
defeatism too readily.  We begin from a state where 
risk bounds of $10^{-6}$ are used quite widely, for 
instance in the solid nuclear waste disposal guidelines
cited above.    It does not seem obvious to me that, with careful
attention to the problems, we could not ensure that catastrophe risks
associated to specific mechanisms are many further orders of
magnitude smaller.  On the contrary, it seems quite clear that in
some cases catastrophe risk
bounds could be substantially reduced.  The RHIC experiments are an
excellent example: had the problem of reducing the risk bounds been
taken seriously, further theoretical research, perhaps combined with
a tentative experimental programme aimed at carefully
testing our understanding of the new physics involved before
running the full experiment, could almost certainly have 
reduced the bounds very significantly. 
 
Of course, this is not to say that risk avoidance is cost free.
One has to accept that more stringent catastrophe risk criteria might
indeed delay or preclude at least some interesting future experiments.
It seems to me we just have to accept this as a fact of life.
One cannot defensibly adopt a 
mindset which requires that every interesting experiment must
be carried out, and that sees every risk analysis as an exercise
in justifying this foregone conclusion.  Human life, collectively
as well as individually, is, after all, fragile.  
Our understanding of nature is
limited, and there are surely many dangers we have not yet
appreciated.  Due caution is appropriate.  

We should not, in any case, rely on speculation about the implications of 
a more cautious catastrophe risk policy.   If it were to 
become clear that it would be effectively impossible to apply a 
policy on catastrophe risks consistently, obviously that policy
would need to be reconsidered.  Unless and until carefully
justified arguments are made, identifying specific examples of
problematic catastrophe risks, it seems premature to worry. 

\item

The justifications given above for risk criteria may strike 
some as a bad policy guide, since they assume that
preserving human lives is in some sense a primary value against
which our actions should be judged. 
Actually, of course, we are guided by many other values.  
Few people consistently act so as to maximise their own life 
expectation, for example: many risky pleasures are widely
indulged in.  
Perhaps we should accept that what applies to us as individuals
applies also to us as a species: worrying about very small risks 
detracts too much from the quality of our existence to be the
best course.  

This is certainly arguable.  On the other hand, current risk policy
tends to count the cost in human lives for a good reason: because that
particular value seems to be more widely shared and more strongly held
than most.  It cannot possibly adequately represent the variety of
individual values we bring to any policy debate, but it is a measure
which, by general consensus, is very important.  Making a generally
acceptable policy for extinction risks on some other basis would
require establishing a fairly firm consensus on what that basis should
be.  No such consensus seems to exist at the moment.

\item

There is what might be termed the argument of dominant risk. 
We face many other extinction risks, some natural
(large asteroid impact), some wholly or partly 
self-created (global nuclear war, catastrophic extinction of species
as a result of human impact on the global ecosystem, catastrophic
climate change as a result of human impact on the global environment).
There is a view
which suggests that a new artificial risk is acceptable if it is 
smaller than existing risks.   A refinement of this view is that 
a new artificial risk is acceptable only if smaller than presently 
unavoidable {\it natural} risks.  In two further common variants of 
these two views, ``smaller than'' is replaced by ``very small compared to''.  

Large asteroid impact seems to be the greatest known 
natural extinction risk that can be reasonably well estimated. 
The risk of the Earth being hit by an asteroid of diameter 
$10$ km is estimated to be $10^{-8}$ per year\cite{chapmanmorrison}.
Such an impact would be so devastating that it is 
generally thought very likely that it would cause mass extinctions
of species, and very plausible that we would be among the species
extinguished.  Accepting that last hypothesis, perhaps at the
price of another order of magnitude, gives an estimate of
$10^{-8}$-$10^{-9}$ per year for this natural extinction risk.
Following the argument of dominant risk leads to the
so-called ``asteroid test'', according to which an artificial 
extinction risk is acceptable if smaller than $\approx 10^{-9}$ per 
year, or 
in the more conservative version, very small compared to $10^{-9}$ per
year.\footnote{Versions of the ``asteroid test'' have been 
discussed as possible justifications for the 
acceptability of the BJSW and DDH risk bounds by several people 
involved in policy formation at CERN and Brookhaven: for example
in W. Pratt et al., Brookhaven National Laboratory Memo to 
T. Ludlam and J. Marburger, 17.2.00.  The test is also considered
in Ref. \cite{calogero}.}

My impression from discussions is that many thoughtful people find 
some version of the argument of dominant risk reasonable, 
but that many equally thoughtful people find this line
of argument entirely irrational.  My sympathies are with the latter.
Why should the existence of one risk, which may be
distressingly high, justify taking another easily 
avoidable risk, which, even if much lower, may still 
be unacceptably high?  Unavoidable natural risks are 
not normally believed to justify wilfully inflicting
avoidable risks on third parties.  
Everyone now living is very likely to die within
the next $120$ years, and would be very likely to die
of natural causes in that timespan even if exposed to no
other risks.  An industry which added slightly to the
natural risk level, annually killing $10000$ people who
had made no choice to accept the extra risk, would not
find much sympathy for the defense that these extra deaths were more or
less lost in the noise compared to natural wastage.  

A further problem specific to the asteroid test is that it assumes that
asteroid extinction risks are either unavoidable or else, though
avoidable in principle, small enough to be tolerable.  In fact,
the asteroid threat is not unavoidable with current and foreseeable technology,
and passive and active counter-measures are being seriously considered.   

That said, let me reiterate that many people seem to be persuaded 
by some version of the argument of dominant risk.  It no doubt
deserves a more careful discussion than is given here.  
The above brief sketch of a counterargument is not meant to 
dismiss the ``asteroid test'' and related criteria out of
hand, but rather just to note that there {\it are} serious
counterarguments.   I do not believe these criteria represent
anything approaching a consensus view.  Unless and until it becomes 
clear that they do, they cannot legitimately be used to
justify catastrophe risks.     

\end{itemize}

\section{Final comment}

The particular artificial extinction risk considered in this paper
is hypothetical, and there are good arguments to suggest that the
actual risk is small or zero.  But, as already noted, we face other
undoubtedly real and not necessarily small artificial extinction
risks.  The arguments above, which suggest that the true costs of 
extinction are generally underestimated, obviously apply generally. 

For instance, while the serious costs associated with artificially
induced global warming are widely (albeit not widely enough) 
appreciated, the extra cost associated with the small risk of a 
truly catastrophic climate change does not seem to have been much 
considered.  Yet it might be that, with proper accounting, the cost of
the risk of climatic catastrophe would be the greater.  
Similarly, although some (insufficient) attention is being paid
to the costs associated with the loss of biodiversity caused by
human impact on the environment, the cost of the risk of a 
catastrophic collapse of the global ecosystem seems to have been
generally neglected.   

In these and other areas where modelling is possible, the arguments
above suggest we should encourage and pay attention to research
into unlikely but not inconceivable catastrophic outcomes, and
try to quantify the risk they represent, rather than focussing 
only on likelier outcomes which may 
be very deleterious but are not truly catastrophic.

\vskip5pt 
\leftline{\bf Acknowledgements} 

Many friends, colleagues and experts in other fields have 
taken considerable time and trouble to help with the 
preparation of this article. 
I would particularly like to thank Clark Chapman for a 
great deal of helpful advice on scientific and
risk policy questions and for many helpful suggestions.  
I am also very grateful to Andy Baker, for supplying
references on risk assessments for nuclear waste storage;
Hans Bethe and Kurt Gottfried, for clarifications of the
history of work at Los Alamos on the atmospheric and oceanic
ignition problems; Guido Altarelli, John Ellis, Jean-Pierre Revol,
Jurgen Schukraft and Gavin Salam for suggesting relevant questions; 
Conor Houghton and Katinka Ridderbos for very helpful criticisms; 
Richard Binzel, Hans Rickman and David Morrison 
for helpful explanations of the risks associated 
with meteor impact and of research on public perception of risk; 
Holger Bech Nielsen, for several interesting discussions on 
collider risks; Francesco Calogero, for helpful discussions and
criticisms and for kindly supplying a draft copy of
Ref. \cite{calogero}. 

It is also a pleasure to thank Stephen Adler, Michael Atiyah,
Guido Bacciagaluppi, David Bailin, Jon Barrett, John Barrow, 
Dorothy Bishop, James Blodgett, Charlotte Bonardi, Harvey Brown, 
Jeremy Butterfield, Joanne Cohn, Jane Cox, Ian Crotty, Matthias
D\"orrzapf, Fay Dowker, Ian Drummond, Michael Froomkin, 
Nicolas Gisin, Peter Goddard, 
Stephen Gratton, Lucien Hardy, Julia Hawkins, 
Ron Horgan, Tobias Hurth, Simon Judge, Anjali Kumar, Peter Landshoff,
Nathan Lepora, Karen McDonald, Jane MacGibbon, David Mermin, 
Hugh Osborn, Sandu Popescu, Patrick Rabbitt, Martin Rees, 
Stefan Reimoser, Peter Ruback, Paul Saffin, Hugh Shanahan, 
Graham Shore, Tony Sudbery,
John Taylor, Daniel Waldram, Peter West and Toby Wiseman
for helpful discussions or correspondence. 

I would like to emphasise that those thanked do not 
necessarily subscribe to the analyses above 
or share any of the views expressed.  

I thank CERN for financial and other support during this work, which 
was also supported by a Royal Society University Research
Fellowship and by the UK Particle Physics and Astronomy Research 
Council. 

{\bf Note added July 2015} \qquad The previous arxiv versions and
the published journal version of this article misstated the
most conservative of BJSW's implied bounds (i.e. those consistent
with the highest probability of catastrophe), multiplying the
probability by a factor of two.

As well as correcting this transcription error, I have taken
the opportunity to add $\approx$ symbols to the relevant figures
in the paragraph stating BJSW's various implied bounds.  There are
two reasons for this.   First, BJSW's calculations produce
rounded rather than precise figures.  Second, precisely 
which catastrophe risk bounds one derives from BJSW's figures
depends on precisely what sort of probabilistic reasoning 
one uses in considering the Moon's survival or destruction
as a result of cosmic ray impacts and the Earth's survival
or destruction as a result of the RHIC experiments.  
For example, should we derive risk bounds for RHIC in which we
have $95 \%$ confidence, given that the Moon has survived?  
Or $99 \%$ confidence?   Or should we use some other approach?
BJSW offer no concrete proposal here.      

It should however be stressed that, whatever approach is used, 
this last point does not allow much scope for stretching 
the bounds.  
The stated risk bounds are derived from 
BJSW's figures if one takes the probability of a dangerous
strangelet to be such that the expected number created over
the lifetime of the Moon is one, which implies a probability
that the Moon survived to be (to very good approximation)
$e^{-1} = 0.3678 \ldots$.   So, the stated risk bounds are 
consistent with an appreciable probability of the Moon having
survived.  In other words, the survival of the Moon does not
give any clear evidence against a catastrophe risk at the level
of the stated bounds.   If we believe we should reject
hypotheses about strangelet creation only if they make the survival of the
Moon seem pretty improbable, then the stated risk bounds are
justified.   Indeed, if we followed one standard method and 
rejected hypotheses about strangelet creation only if the Moon's
survival makes us at least (say) $95\%$ confident that they are incorrect,
we would justify even weaker bounds (consistent with still higher
catastrophe risks).   In any case, BJSW's figures
are consistent with a joint probability of both the Moon
surviving to date and a hypothetical catastrophe at RHIC of $e^{-1}$ times the
stated catastrophe risk bounds.   Since the probability of
hypothetical catastrophe at RHIC is at least as large as this joint probability, it follows that 
(however one treats the separate events of Moon survival and 
hypothetical RHIC catastrophe) no 
catastrophe risk bound better than $e^{-1}$ times the stated figures can be 
derived from BJSW's calculations without introducing further assumptions
that BJSW did not suggest in this part of their discussion. 
In particular, on BJSW's most conservative assumptions,
no risk bound better than $1$ in $36788$ can be derived.   

The omitted factor of $1/2$ to BJSW's most conservative implied risk
bound probabilities makes no
material difference to the arguments of this paper.  
Nor would a further factor of up to $e^{-1}$ to any of BJSW's
implied risk bound probabilities do so.   It is highly debatable
whether any such factor should be included: to the best of my
knowledge, no one has argued that
it should.
As noted above, one can
as well argue that risk bounds should be derived by excluding only
risk probabilities that make the Moon's survival unlikely, in which
case BJSW's risk bounds should be higher rather than lower.     
In any case, given the magnitudes of
the other relevant figures in the argument, the possibility of an
extra factor of $e^{-1}$ seems entirely moot.  
These corrections are made simply for the sake of accuracy.  

A further small correction is that the risk bounds quoted from BJSW
\cite{bjswone,bjswtwo} corresponding to
their Cases I-III were previously listed in order of numerical
size rather than in order of case number.   To remove any 
possible confusion, they are now listed above in case number order, 
so that for example the figures 
$p_{\rm catastrophe} < \approx 2 \times 10^{-10}$, 
$p_{\rm catastrophe} < \approx 10^{-4}$, 
$p_{\rm catastrophe} < \approx 2 \times 10^{-5}$ above 
correspond respectively to Case I, Case II, Case III 
of Ref \cite{bjswone}.    

I am very grateful to Eric Johnson for querying the previously
stated figures and thus drawing my attention to the need for
correction, and for several other very helpful comments and suggestions. 

Let me take the opportunity to add one other comment.   As noted
above, BJSW's initial comments, characterising a risk of
$1$ in $\approx 10000$ of destroying the Earth as ``a comfortable
margin of error'' \cite{bjswone}, are obviously 
inapplicable, and seem to reflect
some surprising confusion on their part.  
Rereading their paper again prompted a guess at the nature of that
confusion.   BJSW's arguments take the following form

{\it If there were a 
high probability (close to one) 
of strangelet-induced catastrophe at RHIC, then 
the expected number of dangerous strangelets produced by cosmic 
rays impacting the Moon over its lifetime would be very high 
($10^4$ or higher).} 

Indeed, and it follows that the probability of the Moon having 
survived would be astronomically low (roughly $e^{-10^4}$ or lower). 
This looks superficially enormously reassuring, given that the Moon
has survived.   The problem 
is that, to argue for the safety of RHIC, we need to exclude not
just a high probability of hypothetical catastrophe but also a small but 
significant probability of hypothetical catastrophe.    It might seem that 
the number $e^{-10^4}$ is so small that rescaling it by any 
significant probability will make no real difference.
Unfortunately -- and I wonder if this may be the point BJSW overlooked
-- the relevant rescaling is in the exponent. 
As noted above, on BJSW's most conservative calculations,
a probability of $10^{-4}$ of hypothetical catastrophe is 
consistent with a probability of $e^{-1}$ of the Moon surviving.



\begin{thebibliography}{99} 
\bibitem{bjswone} 
W. Busza et al., {\it Review of Speculative 'Disaster Scenarios' 
at RHIC}, archived at www.arxiv.org/abs/hep-ph/9910333v1.    
\bibitem{ddh}
A. Dar et al., {\it Will relativistic heavy ion colliders destroy our
planet?}, Phys. Lett. B, 470 142-148 (1999); archived at 
hep-ph/9910471.
\bibitem{kmt}
E. Konopinski, C. Marvin and E. Teller, {\it Ignition of the
Atmosphere with Nuclear Bombs},  Los Alamos Laboratory report
LA-602.  When the present paper was first drafted, 
this reference was archived and freely accessible at
http://lib-www.lanl.gov/la-pubs/00329010.pdf. 
According to the Los Alamos National Laboratory library, access
is presently not permitted, following a directive from the
National Nuclear Security Administration.  I assume this is 
a consequence of heightened security concerns since 11.9.01.  
\bibitem{buck} 
Pearl Buck, {\it The bomb --- the end of the world?}, American Weekly,
March 8, 1959, pp 8-11.  
\bibitem{bethe} 
I am most grateful to Hans Bethe and Kurt Gottfried for this email 
correspondence.
\bibitem{hutrees}
P. Hut and M. Rees, {\it How stable is our vacuum?}, 
Nature {\bf 302}, 508-509 (1983). 
\bibitem{hut} 
P. Hut, {\it Is it safe to disturb the vacuum?}, 
Nucl. Phys. A {\bf 418}, 301c-311c (1984). 
\bibitem{bjswtwo}
R. Jaffe et al., {\it Review of Speculative 'Disaster Scenarios' 
at RHIC}, archived at www.arxiv.org/abs/hep-ph/9910333v2.  
Further minor alterations were made for the final published version:
Rev. Mod. Phys. {\bf 72}, 1125-1140 (2000), 
archived at www.arxiv.org/abs/hep-ph/9910333v3,
\bibitem{concern}
See for example {\it Black Holes at Brookhaven?}, letter to the Editors,
Scientific American, July 1999, p.5; 
{\it Big Bang machine could destroy Earth}, 
article in The Sunday Times (London),
18.7.99, p.26. 
\bibitem{sciamreassurance}
{\it Apocalypse Deferred}, Scientific American, December 1999.  
\bibitem{nsreassurance}
{\it Black Hole Ate My Planet}, New Scientist, 28 August 1999, pp. 24-27. 
\bibitem{bnlweb} 
{\it Committee Report on Speculative ``Disaster Scenarios'' at RHIC:
synopsis}, J. Marburger, 6.10.99. Archived for some while at 
http://www.bnl.gov/bnlweb/rhicreport.html.  
\bibitem{glashowwilson}
S. Glashow and R. Wilson, {\it Taking serious risks seriously},
Nature {\bf 402}, 596-597 (1999). 
\bibitem{akrhicphys}
A. Kent, {\it Problems with empirical bounds for strangelet production
at RHIC}, hep-ph/0009130. 
\bibitem{drns} 
{\it Gambling with the Earth}, New Scientist 7.10.00, p. 4. 
\bibitem{nrpb}
{\it Board Statement on Radiological Protection Objectives for
the Land-based Disposal of Solid Radioactive Wastes},
Documents of the NRPB {\bf 3}, 3 (1992).  Abstract available at
http://www.nrpb.org.uk/Absd3-3.htm
\bibitem{chapmanmorrison}
C. Chapman and D. Morrison, {\it Impacts on the Earth by asteroids
and comets: assessing the hazard}, Nature {\bf 367}, 33-40 (1994).
\bibitem{calogero} 
F. Calogero, {\it Might a laboratory experiment now being planned destroy
planet Earth?}, Interdisciplinary Science Reviews {\bf 25}, 191-202 (2000).
\end{thebibliography}
\end{document}